\documentclass[
journal=ancac3,
manuscript=article,
]{achemso}
\SectionNumbersOff

\usepackage[UKenglish]{babel}

\usepackage{amsfonts}
\usepackage{amsmath}
\usepackage{amssymb}
\usepackage{upgreek}
\usepackage{dsfont}
\usepackage{siunitx}
\usepackage{esint} 
\usepackage{mathrsfs} 

\usepackage{accents}

\usepackage{physics}
\usepackage{siunitx}
\DeclareSIUnit{\litre}{l} 

\usepackage{hyperref}

\usepackage{graphicx} 
\usepackage{overpic} 

\usepackage[font=sf,labelfont=bf]{caption}
\usepackage{enumitem}

\usepackage[dvipsnames]{xcolor}
\definecolor{ts_BG}{HTML}{1D1F21}
\definecolor{ts_k}{HTML}{282A2E}
\definecolor{ts_r}{HTML}{A54242}
\definecolor{ts_g}{HTML}{8C9440}
\definecolor{ts_y}{HTML}{DE935F}
\definecolor{ts_b}{HTML}{5F819D}
\definecolor{ts_m}{HTML}{85678F}
\definecolor{ts_c}{HTML}{5E8D87}
\definecolor{ts_gray}{HTML}{707880}

\usepackage{float}

\title{Fourier Plane Tomographic Spectroscopy Reveals Orientation-Dependent Multipolar Plasmon Modes in Micrometer-Scale Janus Particles}
\author{Felix H. Patzschke}
\author{Frank Cichos}
    \email{cichos@physik.uni-leipzig.de}
    \affiliation{Molecular Nanophotonics Group, Peter Debye Institute for Soft Matter Physics, Leipzig University, 04103 Leipzig, Germany}

\abbreviations{Au, AuNP, BFP, BFP, LMI, LSP, LSPR, pJP, PS, SP}

\begin{document}

\begin{abstract}
Plasmonic Janus particles, comprising dielectric cores with thin metallic caps, exhibit complex optical properties due to their asymmetric structure. Despite applications in active matter research, their orientation-dependent scattering properties remain largely unexplored. We introduce Fourier plane tomographic spectroscopy for simultaneous four-dimensional characterization of scattering from individual micrometer-scale particles across wavelength, incident angle, scattering angle, and polarization. Combining measurements with finite-element simulations, we identify discrete spectral markers in visible and near-infrared regions that evolve predictably with cap orientation. Spherical-harmonics decomposition reveals these markers arise from three distinct multipolar modes up to fifth order: axial-propagating transverse-electric, transverse-propagating transverse-electric, and transverse-propagating axial-electric, with retardation-induced splitting. We observe progressive red-shifts and linewidth narrowing of higher-order resonances, demonstrating curvature's influence on mode dispersion. Orientation-specific scattering patterns exhibit polarization-dependent features enabling optical tracking of particle rotation. Our framework applies to diverse material combinations and geometries, offering a toolkit for designing orientation-responsive nanoantennas, reconfigurable metasurfaces, and active colloidal systems with tailored light-matter interactions.
\end{abstract}



\section{}

Precise control of light at the nanoscale through plasmonic excitations has emerged as a cornerstone technology enabling advances in next-generation sensing, energy harvesting, and quantum information systems. A significant aspect of these light-matter interactions (LMIs) is the excitation of surface plasmons (SPs) -- collective oscillations of electrons at metal-dielectric interfaces.\cite{barnesSurfacePlasmonSubwavelength2003,MaierPlasmonics,zhang2012surface} When these oscillations are confined to metallic nanostructures, they give rise to localized surface plasmons (LSPs) which generate highly localized enhancements of electromagnetic fields.\cite{MaierPlasmonics,schullerPlasmonicsExtremeLight2010,mcoyiDevelopmentsLocalizedSurface2024,zhang2012surface}
These plasmonic phenomena find immediate application in critical challenges, from ultra-sensitive biosensors for point-of-care medical diagnostics\cite{liPlasmonicbasedPlatformsDiagnosis2019,tokelAdvancesPlasmonicTechnologies2014,jinPlasmonicNanosensorsPointofcare2022} to efficiency-enhanced photovoltaic cells addressing global energy demands,\cite{atwaterPlasmonicsImprovedPhotovoltaic2010,arinzePlasmonicNanoparticleEnhancement2016,wangPlasmonicsMeetsPerovskite2024} and quantum plasmonic devices enabling imaging beyond the diffraction limit.\cite{kawataPlasmonicsNearfieldNanoimaging2009,gramotnevPlasmonicsDiffractionLimit2010,leeQuantumPlasmonicSensors2021}
The sensitivity of localized surface plasmon resonances (LSPRs) to geometry and local refractive index has been extensively studied in various nanostructures, including nanorods,\cite{chaudhari2014spatiotemporal,funstonPlasmonCouplingGold2009} nanotriangles,\cite{shufordMultipolarExcitationTriangular2005,zhangOpticalPropertiesSilver2010,wangPlasmonicEigenmodesIndividual2015,Islam2021} and nanoparticle aggregates.\cite{zuloagaQuantumDescriptionPlasmon2009,gunnarssonInterparticleCouplingEffects2001,g.m.vieiraPlasmonicPropertiesClosePacked2019,fergusson2023plasmonic} These nanostructures act as resonators, supporting a collection of fundamental oscillation modes. The resonance frequencies of these modes depend strongly on the size and shape of the underlying structure.\cite{mayer2011localized,Kelly2003,prodanHybridizationModelPlasmon2003,wangPlasmonicEigenmodesIndividual2015} Most research has focused on plasmonic structures smaller than or comparable to the probing wavelengths, which allows for straightforward excitation and analysis of fundamental LSPRs.\cite{schullerPlasmonicsExtremeLight2010,gramotnevPlasmonicsDiffractionLimit2010,kauranenNonlinearPlasmonics2012}

Meanwhile, higher-order resonances, which provide valuable information about the spatial characteristics of plasmonic oscillation modes, have rarely been studied. While they are known to appear as contributions to plasmonic excitations on various nanostructures,\cite{penninkhofOpticalPropertiesSpherical2008,tanakaNanoscaleInterferencePatterns2012} investigating these higher-order LSPRs in the regime where they constitute principal features, requires larger plasmonic structures, where higher order modes of the fundamental LSP modes can couple to incident visible light fields.\cite{funstonPlasmonCouplingGold2009,vesseurDirectObservationPlasmonic2007,schmidtExploringPlasmonicCoupling2015,ohadSpatiallyResolvedMeasurement2018} Structures with reduced symmetry are particularly interesting, as they split fundamental LSP modes along their principal axes\cite{halas11angle,pakizehUnidirectionalUltracompactOptical2009} and exhibit multipolar resonances that map to geometric features.\cite{youPolarizationDependentMultipolarPlasmon2012} Analysis of the anisotropic optical response of such nanostructures offers unique insights into orientation-dependent LMIs.\cite{pakizehUnidirectionalUltracompactOptical2009,Islam2021,sonnichsenDrasticReductionPlasmon2002,mockDistanceDependentPlasmonResonant2008,mirin2009light,youPolarizationDependentMultipolarPlasmon2012} Micrometre-sized plasmonic Janus particles (pJPs) are ideal candidates for studying fundamental LSPR modes in the optical regime.\cite{vandorpeSemishellsVersatilePlasmonic2011} These particles, consisting of a dielectric core with a thin metallic coating on one hemisphere, bear strong structural resemblance to spherical cap systems, enabling efficient SP coupling\cite{gongSurfacePlasmonCoupling2017} and exhibit multiple LSP modes with orientation-dependent excitations.\cite{halas11angle,halas11transfer} While classical Mie theory\cite{Mie1908} only makes predictions for maximally symmetric particles, it provides a valuable reference framework for the interpretation of findings regarding higher-order plasmonic modes of fundamental LSPRs. The strong plasmonic responses of pJPs have enabled their widespread use in studying self-organization and active matter.\cite{jiang2010active,rey2023light} Theoretical and numerical studies have revealed complex LMIs in these systems, predicting counter-intuitive phenomena such as stable rotation induced by linearly polarized light fields.\cite{Ilic2017,BA}

Despite their widespread use in active matter research, the orientation-dependent plasmonic responses of individual micrometer-scale Janus particles remain incompletely characterized, limiting their potential for precision optical manipulation and sensing applications. Here, we address this critical knowledge gap by developing Fourier plane tomographic spectroscopy (building on back focal plane principles) to provide comprehensive four-dimensional characterization of individual pJP light-matter interactions.\cite{MONA-BFP-photonic-crystals,MONA-BFP-photonic-stop-bands,ohadSpatiallyResolvedMeasurement2018} In this study, we investigate plasmonic Janus particles and map the intensity of scattered light resolved for wavelength and scattering angle while varying the direction of illumination. This approach reveals the contributions of higher-order excitations of specific SP modes to the optical responses. Complementing our experimental work, we perform numerical simulations that show excellent agreement with measurements, allowing us to correlate peaks in the scattering spectra to orientation-dependent surface plasmon modes.

This quantitative understanding of orientation-dependent plasmonic responses provides essential design principles for emerging applications in active metamaterials, optical manipulation systems, and next-generation sensing platforms where precise control over light-matter interactions is paramount. Through this combined approach, we gain new insights into the LMIs of large, anisotropic, plasmonic structures in an intermediate regime between two well-understood limits -- the dipole resonances of small particles, where incident optical fields are effectively uniform,\cite{cox2002experiment,GouesbetGrehan} and the diverse SP excitations on extended thin films\cite{economou1969surfacePlasmons}. This comprehensive framework enables rational design of orientation-responsive plasmonic devices and provides a pathway toward next-generation metamaterials with dynamically tunable optical properties.

\section{Results}

We investigated the light-matter interactions of plasmonic Janus particles using a spectroscopic technique developed for this purpose: Fourier Plane Tomographic Spectroscopy. This method combines selective dark-field illumination with back focal plane (BFP) spectroscopy, enabling precise control of illumination direction while simultaneously resolving scattered light intensity as a function of both wavelength and scattering angle. This approach provides comprehensive four-dimensional characterization of asymmetric particles' optical response, capturing the full angular and spectral distribution of scattered light under controlled illumination conditions.

\begin{figure*}[htbp]
    \includegraphics[width=\textwidth]{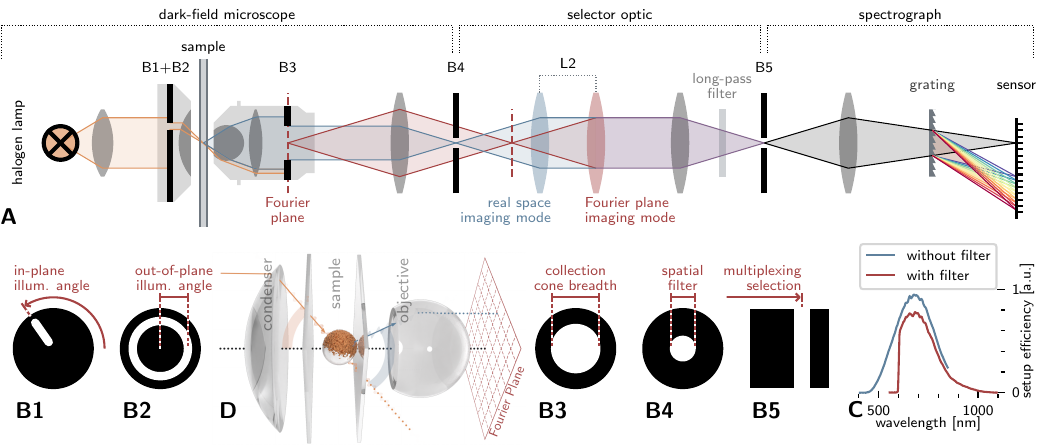}
    \caption{
    \textbf{A:} Schematics of the imaging light path.
    \textbf{B1-B5:} Schematics of the apertures.
    {\sffamily B1-B3} lay in Fourier planes, {\sffamily B4} lies in an image plane. Depending on the placement of the lens {\sffamily L2}, either the image plane or the Fourier plane may be imaged onto {\sffamily B5}, and subsequently the camera sensor.
    \textbf{C:} Spectral response function of the setup. Measurements with and without the long-pass filter were spliced together to obtain spectra covering the entire sensitivity range.
    \textbf{D:} In the immediate vicinity of the particle under observation, the ambient refractive index is virtually homogenous, due to the usage of immersion oil inside the sample.
    }
    \label{fig:setup}
\end{figure*}

\paragraph*{Fourier Plane Tomographic Spectroscopy} Angle-resolved scattering spectra were acquired utilizing a custom-designed optical configuration, depicted in Fig. \ref{fig:setup}A, constructed around a standard dark-field microscope. Precision apertures positioned in the illumination pathway (B1 and B2) constrained the incident illumination to within $\SI{0.096}{\steradian}$ solid angle along predetermined directions. The instrument's optical path was configurable to either directly image the sample plane or project the back focal plane (B3) of the objective lens onto an intermediate imaging plane. A spatial filter (B4) positioned at an intermediate image plane isolated scattered light exclusively from the particle under investigation, and spectral dispersion was accomplished via a transmission grating. Full details of the optical configuration, sample preparation, and experimental procedures are provided in the Methods section.

Real-space imaging facilitated the selection of individual particles and spectral measurements without angular resolution, while back focal plane (BFP) imaging provided spectrally resolved Fourier-space scattering distributions. To acquire these datasets, the slit B5 was systematically translated across the BFP image during the recording process, enabling the camera to sequentially capture vertical lines of spectrally-dispersed scattering information.

For spectral calibration, the objective's back aperture (B3) was fully opened to its maximum numerical aperture of 1.3, allowing direct transmission of light from the dark-field illumination pathway into the imaging system. The resulting spectral efficiency curves of the optical configuration are presented in Figure \ref{fig:setup}C. The spectral range of the measurements was constrained at shorter wavelengths by the emission spectrum of the light source, while the upper wavelength limit was determined by the quantum efficiency of the sCMOS detector and partial absorption of near-infrared radiation by the optical components in the beam path.

Samples comprised immobilized plasmonic particles sandwiched between two glass cover slides and embedded in immersion oil. This sample structure was chosen in order to provide a homogenous refractive index in the vicinity of the probed particles, as illustrated in Figure \ref{fig:setup}D, minimizing perturbations through boundary effects such as thin-film interference.

\begin{figure*}
\centering
    \includegraphics[width=\textwidth]{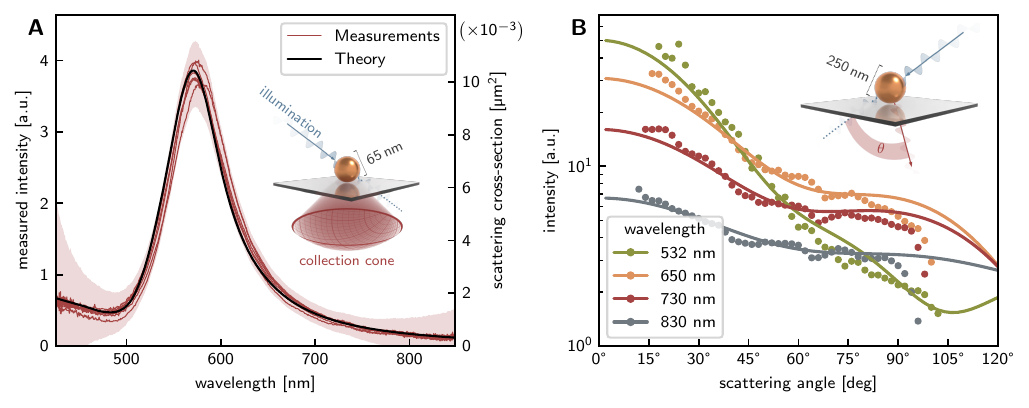}
    \caption{Validation measurements.
    {\sffamily\bfseries A:}
    In an angle-accumulated measurement, a cone of scattered light is collected (see inset) and spectrally analyzed at once.
    For a \mbox{65 nm} AuNP, the theoretical scattering cross-section is well-approximated by this, falling within a standard deviation (shaded area) of the measurements.
    {\sffamily\bfseries B:}
    In an angle-resolved measurement (see inset), the light emitted under a specific scattering angle $\theta$ is isolated before spectral dispersion.
    For a spherical AuNP \mbox{(d = 250 nm)}, the measured angular intensity profiles (points in diagram) match the predictions of Mie theory\cite{Mie1908} (lines in diagram) for various wavelengths.
    All measured intensities were scaled by the same constant factor to match the theory curves.
    This demonstrates the accurate capture of the system's response in both the $\theta$- and the $\lambda$-dimension.
    }
    \label{fig:AuNP}
\end{figure*}

\paragraph*{Single Gold Nanoparticle Spectra}
To validate the spectroscopic performance of our experimental setup, we conducted measurements on gold nanoparticles (AuNPs) with a diameter of $65\, \si{\nano\metre}$. Fig. \ref{fig:AuNP}A demonstrates the excellent agreement between our measured scattering spectra and theoretical predictions based on Mie theory,\cite{Mie1908} utilizing the complex refractive index of gold reported by Johnson and Christy.\cite{Johnson1972} Statistical analysis of the measurements revealed minor variations in peak positions ($\SI{2.2}{\nano\meter}$ standard deviation) attributable to the particle size distribution, alongside a systematic red-shift of $\SI{3.6}{\nano\meter}$ relative to theoretical predictions. For particles substantially smaller than the incident wavelength (characterized by a size parameter $x = (2\pi n r)/\lambda \ll 1$), the dipole approximation predicts angular scattering distributions that are invariant with respect to shape, resulting in measured intensities that directly correspond to total scattering cross-sections. In our experimental configuration, the size parameter ranges from 0.34 at $\lambda=\SI{900}{\nano\meter}$ to 0.77 at $\lambda=\SI{400}{\nano\meter}$, which deviates somewhat from the strict dipole approximation regime. This deviation explains the observed red-shift of the measured resonance peak: multipole contributions to the angular distribution of scattered light become progressively less significant at longer wavelengths, resulting in proportionally greater collection efficiency through the objective back aperture in this spectral region.

For angle-resolved measurements, we used larger $\SI{250}{\nano\metre}$ AuNPs. Measured angular scattering patterns match theoretical predictions,\cite{BohrenHuffman, GouesbetGrehan} as we demonstrate in Fig. \ref{fig:AuNP}B. The forward-scattering becomes more pronounced at shorter wavelengths, where the size parameter $x$ increases from 1.2 ($\lambda=\SI{1000}{\nano\metre}$) to 3.0 ($\lambda=\SI{400}{\nano\metre}$).

\begin{figure*}[h]
    \centering
    \includegraphics[width=\textwidth]{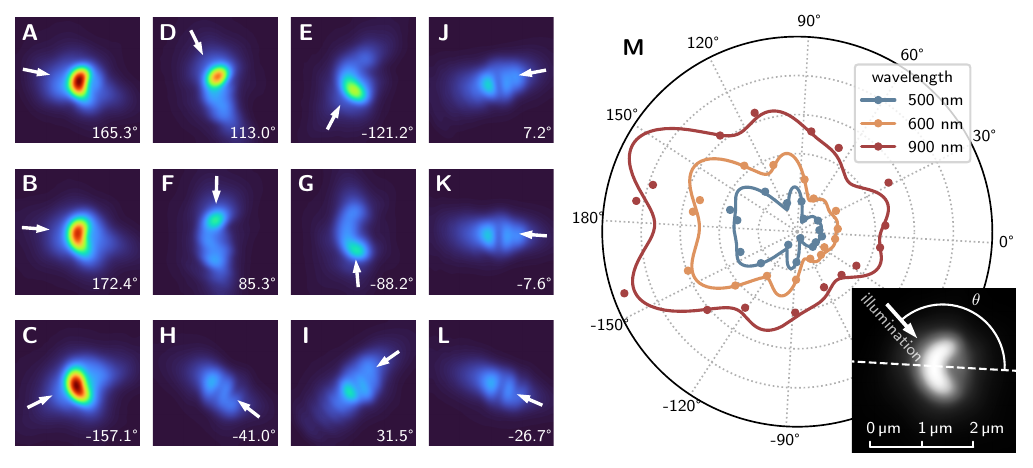}
    \caption{{\bfseries A-L:}
    Dark-field images of a pJP under various selective illumination modes. The respective in-plane illumination angles are indicated by arrows and noted in the lower right corner.
    {\bfseries M:}
    Apparent brightness of the pJP depending on the in-plane illumination angle, for different wavelengths.
    The profile lines were obtained by fitting to a finite Fourier series.
    A standard dark-field image of the same pJP, annotated with its symmetry axis and the definition of the illumination angle (here $\theta$), is given in the lower right corner.
    }
    \label{fig:tomography}
\end{figure*}

\paragraph*{Illumination-Angle-Dependent Scattering Intensity}
Micron-sized pJPs exhibit remarkably complex angle-dependent scattering behavior arising from their structural anisotropy, despite maintaining a spherically symmetric core. Our experimental configuration enables precise manipulation of illumination angles through controlled rotation of aperture B1 within the illumination pathway while preserving fixed particle orientation throughout measurements.

Figures \ref{fig:tomography}A-L present dark-field images of a single pJP under 12 different illumination angles, demonstrating how the light-scattering activity is localized based on the orientation of the plasmonic structure and the  direction of illumination.
When light is incident on the Au cap side (figures \ref{fig:tomography}A-G), the scattered intensity is strongly localized, characteristic of plasmonic scattering from the metallic surface. The PS side exhibits minimal scattering due to its low refractive index contrast with the surrounding medium. Under PS-side illumination (figures \ref{fig:tomography}H-L), scattering is observed from both hemispheres, separated by a distinctive dark region. This pattern can be attributed to plasmon-mediated scattering from the Au cap coupled with subsequent refraction through the PS core.

The apparent brightness of the pJP in dark-field imaging depends systematically on both illumination direction and wavelength. Analysis of spectrally resolved scattering under restricted illumination conditions reveals wavelength-dependent angular profiles (Fig. \ref{fig:tomography}M). The scattering intensity is consistently higher for Au-side illumination across all wavelengths, with the intensity contrast between Au-side and PS-side illumination being most pronounced at shorter wavelengths. The angular profiles at shorter wavelengths display multiple well-defined extrema, while these features become less distinct at longer wavelengths, consistent with the transition between different scattering regimes.

\begin{figure*}[h]
    \centering
    \includegraphics[width=\textwidth]{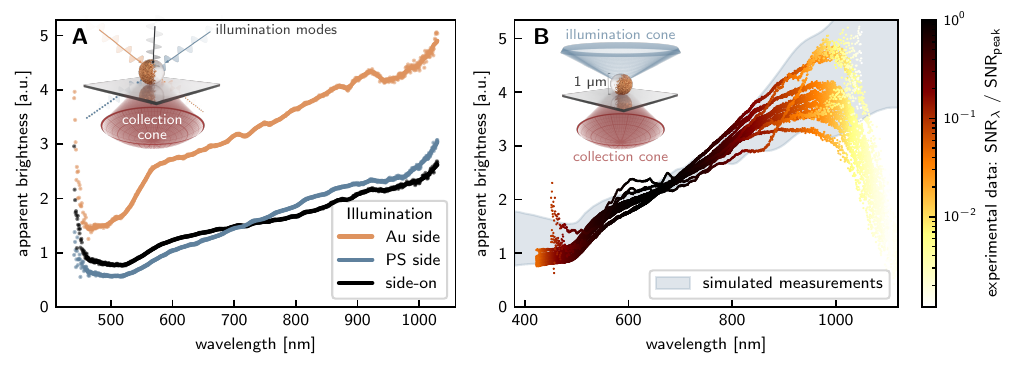}
    \caption{
        Measured scattering spectra of $\SI{1}{\micro\meter}$ pJPs.
        {\sffamily\bfseries A:} Scattering spectra of a single pJP, with different settings of the selective illumination.
        Light was incident from the Au side (yellow curve), from the PS side (blue curve) and side-on (black curve).
        {\sffamily\bfseries B:} Under standard dark-field illumination, light is incident from a range of directions simulateneously.
        The shaded area depicts the range of simulated measurements under the same conditions.
    }
    \label{fig:spectra-measured}
\end{figure*}

\paragraph*{Single Janus Particle Scattering Spectra}
The recorded scattering spectra of individual pJPs under selective illumination from specific directions (Au side, PS side, and perpendicular to the symmetry axis) are shown in Fig. \ref{fig:spectra-measured}A. In the wavelength range from $600$ to $\SI{1000}{\nano\meter}$, the scattering intensity increases monotonically for all illumination geometries. The measured scattering response was consistently highest for illumination from the Au side, as evidenced by the angular profiles. The Au-side illumination spectrum exhibits a distinct spectral shoulder at $\SI{550}{\nano\meter}$. This feature appears diminished under perpendicular illumination and is absent when illuminating from the PS side.

Figure \ref{fig:spectra-measured}B illustrates the collective dark-field scattering spectra acquired from multiple pJPs juxtaposed with corresponding finite-element simulations that incorporate the objective's numerical aperture constraints. Quantitative comparison reveals excellent correlation between experimental measurements and computational predictions, with the experimentally obtained spectral profiles consistently falling within one standard deviation of the mean simulated response throughout the predominant wavelength domain examined. This robust agreement validates our numerical approach and confirms the underlying physical mechanisms governing the plasmonic response of these asymmetric structures.

Both measured and simulated spectra exhibit a consistent monotonic increase in scattering intensity across the $500$-$\SI{1000}{\nano\meter}$ spectral range, with a characteristic shoulder feature at $\SI{550}{\nano\meter}$ evident throughout all experimental observations. This spectral profile differs markedly from homogeneous Au spheres of equivalent dimensions, which characteristically display a pronounced resonance peak followed by a slowly decaying intensity at longer wavelengths.\cite{Mie1908,GouesbetGrehan,Johnson1972} The signature dipolar plasmon resonance typically prominent in smaller AuNPs manifests merely as a subtle shoulder in our spectral data, subsumed by the dominant trend of increasing scattering intensity with wavelength.

At wavelengths exceeding $\SI{900}{\nano\meter}$, the measured spectra exhibit decreased signal-to-noise ratio. This spectral artifact stems predominantly from the inherent detector noise limitations, as the quantum efficiency of our detection system diminishes significantly in this near-infrared region. Notably, while our experimental measurements consistently demonstrate decreasing scattering intensity approaching negligible levels at longer wavelengths, computational simulations predict continued enhancement in scattering efficiency beyond this spectral boundary. This discrepancy between experimental observations and theoretical predictions warrants further investigation into near-infrared plasmonic behavior.

\begin{figure*}
    \centering
    \includegraphics[width=\textwidth]{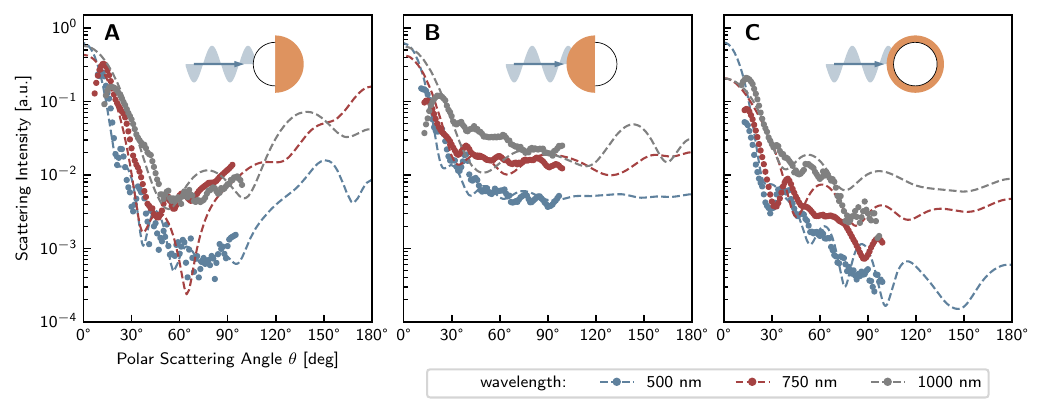}
    \caption{Scattering intensity of the pJP versus scattering angle for various wavelengths.
    The points correspond to measured intensities while the lines are simulation results.
    {\sffamily\bfseries A:} PS side illumination.
    {\sffamily\bfseries B:} Au side illumination.
    {\sffamily\bfseries C:} side-on illumination.
    }
    \label{fig:jp-mieplots-oneline}
\end{figure*}

\paragraph*{Angular Scattering Intensity of Janus Particles}
Fourier Plane Tomographic Spectroscopy provided comprehensive characterization of the angular distribution of scattered light intensity. We conducted measurements under three precise illumination configurations: Au-side axial, PS-side axial, and side-on illumination (illustrated in Fig. \ref{fig:jp-mieplots-oneline} insets). For rigorous quantitative comparison between experimental measurements and computational simulations, we extracted one-dimensional intensity profiles along the polar coordinate from the Fourier plane images. Figure \ref{fig:jp-mieplots-oneline} presents both the measured and simulated scattering profiles.

Consistent with established Mie scattering principles, the scattered light intensity demonstrates pronounced forward ($\theta=0$) directionality across all illumination geometries. However, the angular distributions exhibit distinct qualitative features that remain consistent throughout the analyzed spectral range. Under PS-side illumination (Fig. \ref{fig:jp-mieplots-oneline}A), the forward scattering peak achieves maximum intensity, followed by a characteristic minimum between $\SI{45}{\degree}$ and $\SI{60}{\degree}$ scattering angle, with subsequently increasing intensity at larger angles. For Au-side illumination (Fig. \ref{fig:jp-mieplots-oneline}B), the forward scattering peak exhibits significant broadening, and notably, the intensity stabilizes at a plateau value for intermediate scattering angles rather than continuously decreasing. This broader angular distribution yields higher total scattered intensity when integrated across all angles. Under transverse illumination (Fig. \ref{fig:jp-mieplots-oneline}C), the scattered intensity demonstrates a steep, nearly monotonic decrease with increasing scattering angle. These characteristic features are accurately reproduced in the simulated far-field patterns, providing robust validation of our experimental observations.

\section{Discussion}

\begin{figure*}
    \centering
    \includegraphics[width=\textwidth]{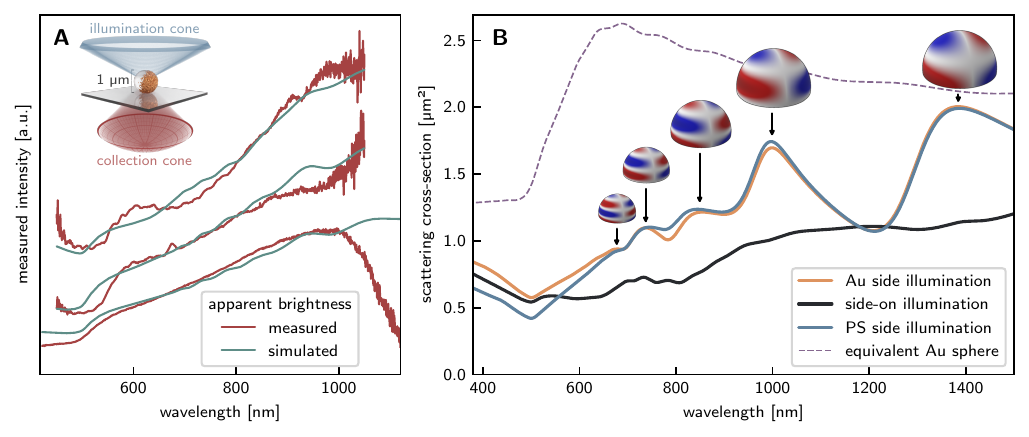}
    \caption{\textbf{A:} A selection of measured dark-field spectra of pJPs alongside the closest-matching simulated spectra. The simulations took into account the combination of incident directions through the dark-field condenser, as well as the finite numerical aperture of the objective.
    \textbf{B:} Simulated scattering spectra of the pJP under illumination from the Au side (yellow), from the PS side (blue) and side-on (black). The dashed line indicates the scattering spectrum of an equivalently sized Au particle. Its less pronounced and more tightly spaced peaks resemble those of the pJP's scattering spectrum under side-on illumination. Peaks in the spectra for axial illumination are labelled with sketches of the associated surface plasmons.
    }
    \label{fig:spectra-principal}
\end{figure*}

\paragraph*{Scattering Spectra and Comparison to Mie Theory}
To complement our experimental findings, we performed finite-element simulations to calculate the far-field scattering intensity patterns of pJPs under different illumination directions. In Figure \ref{fig:spectra-principal}A, we demonstrate that the measured scattering spectra can be reproduced well, using the simulation results.

To that end, the finite numerical aperture of the real optics and the range of illumination angles in the dark-field configuration had to be taken into account. In contrast, the scattering spectra for unidirectional illumination modes and accumulation over all scattering angles are presented in \ref{fig:spectra-principal}B. Although the simulated spectra show some notable differences from the measured dark-field spectra, key features are preserved between both datasets. The general increasing trend in scattering intensity with wavelength is maintained, including the characteristic upturn at $\SI{500}{\nano\meter}$. For both Au-side and PS-side illumination, the simulated scattering cross-sections exhibit wavelength-dependent behavior similar to the measured dark-field spectra, with a minimum at $\SI{500}{\nano\meter}$ followed by monotonic growth at longer wavelengths. However, the simulated spectra reveal distinct resonance peaks that are not readily apparent in the experimental measurements. Under side-on illumination, the simulated scattering cross-section increases more gradually with wavelength, showing less pronounced spectral features except for a cluster of minor peaks between $700$ and $\SI{800}{\nano\meter}$. The presence of less pronounced peaks in closer proximity to each other under this illumination mode indicates the excitation of a more complex LSP mode structure.

The choice theoretical framework for the description of LMIs with a size parameter close to 1 is the theory of Mie, which presents an analytical solution of the electromagnetic wave equation for an incident plane wave and a homogeneous, spherical particle.\cite{Mie1908} However, the lower degree of symmetry of our Janus Particle w.r.t. a similar-sized gold sphere, and coupling between the outer and inner surfaces of the pJP's cap effect a significantly altered plasmonic activity and, consequently, scattering response.

In the scattering spectrum of a solid Au sphere of equivalent size to the pJP we considered, the distinct peaks which we find for axial illumination, are not discernible, as we show in Fig. \ref{fig:spectra-principal}B. This is due to the higher degree of symmetry of the sphere: Where the principal LSP modes are split by orientation for the pJP,\cite{halas11angle} on a sphere, all modes are threefold degenerate and can be excited by light incident from any direction. Consequently, the resonances of many different LSP modes overlap in the scattering spectrum of the gold sphere, as they do in the side-on-illuminated scattering spectra of the pJP.

The angular intensity distributions exhibit the same qualitative behaviour for the pJP as those that one might calculate for a solid Au sphere: As the wavelength increases, the non-global maxima beside the consistently present forward-scattering peak become more well-distinguished in magnitude and fewer in number. The same happens as the direction of illumination is changed from side-on to axial. Both parameter changes can, in the context of Mie theory, be understood as a decreasing size parameter and thus the transition away from the ray optics regime $\left( \lambda \ll R \right)$ and closer to the scattering dipole model $\left( \lambda \gg R \right)$.

\begin{figure*}
    \includegraphics[width=\textwidth]{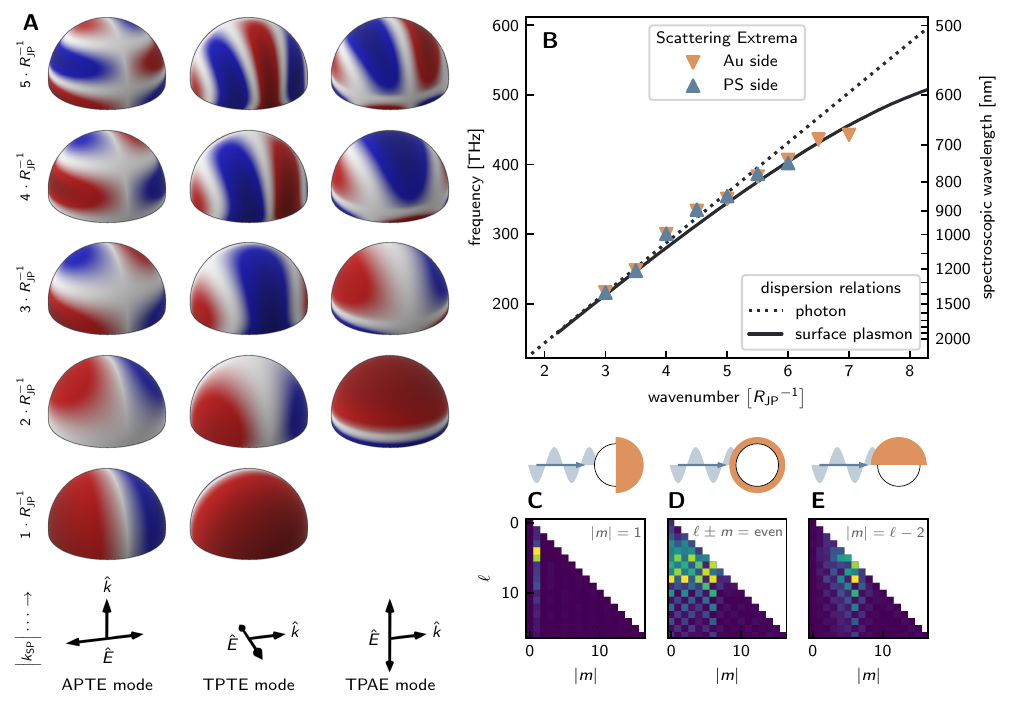}
    \caption{
        \textbf{A:} Sketches of the surface plasmon modes of a hemispherical gold cap.
        The given wavenumber corresponds to the fundamental standing wave.
        The arrows at the bottom indicate the orientation of the outside light field.
        The bottommost modes of each column are those that would be excited in a spatially invariant external electric field.
        \textbf{B:} Excitation wavelengths of peaks and valleys in the axial illumination scattering spectra vs. spatial frequency of the electric field on the surface of the Au cap. The inferred wavenumber is given in units of inverse pJP radii, $\SI{505}{\nano\meter}$.
        \textbf{C-E:} Decompositions of the electric field on the cap's surface for an excitation wavelength of $\SI {892}{\nano\meter}$ into spherical harmonics. The heatmaps visualize the amplitudes of each component $Y_\ell^m$ of the expansions. In \textbf{C}, the orientation promotes APTE mode excitation. The non-zero contributions fulfill $\vert m \vert = 1$. \textbf{D} and \textbf{E} correspond to the TPTE and TPAE modes, with the significant contributions matching the selection rules $\ell \pm m=\text{even}$ and $\vert m \vert = \ell-2$, respectively.
    }
    \label{fig:basic-modes}
\end{figure*}

\paragraph*{Surface Plasmon Modes of the Gold Cap}
The scattering properties of metallic nanostructures arise from the excitation of localized surface plasmons on their surfaces.\cite{amendola2017surface}
Previous studies on smaller structures of similar geometry have shown that in the long-wavelength limit where $k \ll R_\mathrm{JP}^{-1}$, the response is dominated by two dipole modes: transverse-electric (TE) and axial-electric (AE).
The TE mode is twice degenerate due to the rotational symmetry of the cap about its axis.\cite{halas11angle}

Here, the larger size of the particle enables the excitation of higher-order LSP modes on the gold cap. To analyze these modes quantitatively, we develop a comprehensive mapping of the hemispherical cap geometry to the unit sphere $\mathcal{S}^2$. This mapping is constructed by projecting the outer (gold-oil interface) and inner (gold-polystyrene interface) surfaces of the cap onto the complete unit sphere $\mathcal{S}^2$, with the outer surface corresponding to the upper hemisphere ($ \left[0, \pi/2\right] \ni \theta \mapsto \theta$) and the inner surface to the lower hemisphere ($\left[\pi/2, \pi\right] \ni \theta \mapsto \pi-\theta \in \left[0, \pi/2\right]$). Therein, the boundaries of either sub-domain at $\theta=\pi/2$, representing the physical rim where the gold film terminates, are identified.

This spherical mapping shows constructively, that the domain of the LSP modes is topologically equivalent to the unit sphere and thus enables us to express the fundamental spatial oscillations of the surface charge density in terms of spherical harmonics $Y_\ell^m (\theta,\phi)$, where $\theta$ and $\phi$ are spherical coordinates parametrising $\mathcal{S}^2$. For resonant surface plasmon modes, this topology leads to a quantization of allowed wave numbers, approximated by $$k_\mathrm{SP} = \frac{\ell}{R_\mathrm{JP}}\ ,$$ where $R_\mathrm{JP}$ is the effective radius of the gold cap. This relation describes standing wave resonances on the cap surface.

While the multipole order $\ell$ determines the total spatial frequency of the oscillation, $m$ represents the number of complete oscillations of $Y_\ell^m (\theta,\phi)$ along a closed azimuthal curve at fixed $\theta$, as evident from the analytical form of the spherical harmonics.\cite{Bronstein2013} Along a meridional path (fixed $\phi$), the number of oscillations is given by $\ell-\vert m \vert$. Therefore, the pair $\left( \ell-\vert m \vert , m \right)$ can be interpreted as the components of the surface plasmon wave vector in polar and azimuthal directions, respectively.

Though the cap's surface is topologically equivalent to that of a spherical particle, its hemispherical geometry splits the allowed oscillations into three distinct surface plasmon modes, characterized by the direction of charge density oscillation and the direction of plasmon propagation. The oscillation direction is determined by the polarization of the incident electric field, while the propagation direction is set by its wave vector. We therefore identify:

i) the axially-propagating, transverse-electric (APTE) mode (Figure \ref{fig:basic-modes}A, left column): This mode is excited when the electric field is polarized perpendicular to the particle's symmetry axis. Due to cylindrical symmetry around this axis, the surface charge distribution must transform as a dipole under rotation—maintaining one complete oscillation around any circle of fixed $\theta$. This requirement arises from the transverse dipolar nature of the excitation field coupling to the surface plasmon, which necessitates $m = \pm1$ as the only possible values that preserve the dipolar field symmetry while allowing propagation along the axis. This mode is doubly degenerate due to rotational symmetry about the particle axis.

ii) the transverse-propagating, transverse-electric (TPTE) mode (fig. \ref{fig:basic-modes}A middle): For the TPTE mode, the surface charge density must have odd symmetry about the plane orthogonal to the polarization direction. The coupling between the inner and outer hemispheres requires that, on the entire unit sphere, there must be an even number of closed azimuthal curves which are nodes of the standing wave. Consequently, we determine that the major contributions to the TPTE surface charge oscillation must be spherical harmonics $Y_\ell^m$ where $\ell \pm m$ is even and $\ell>0$.

iii) the transverse-propagating, axial-electric (TPAE) mode (fig. \ref{fig:basic-modes}A right): This mode is excited when the electric field is polarized parallel to the rotational symmetry axis of the pJP. The surface charge density distribution exhibits a specific constraint due to coupling between the inner and outer surfaces of the gold cap. Specifically, at the poles, the surface charge density must maintain the same sign on both the inside and outside surfaces; otherwise, charge would flow through the volume, constituting a volume plasmon that requires significantly higher excitation frequencies. This constraint necessitates the presence of a nodal curve on each hemisphere that encircles the pole without intersecting the rim. Spherical harmonics $Y_\ell^m$ with $\vert m\vert = \ell-2$ satisfy this condition, and we therefore identify these as the fundamental representation of the TPAE mode.

From these constraints on the LSP modes, it follows that a surface plasmon resonance with $k_\mathrm{SP}=1 \cdot {R_\mathrm{JP}}^{-1}$ exists only in the TPTE and APTE modes, but not in the TPAE mode, where the minimum resonance wave number is $2 \cdot {R_\mathrm{JP}}^{-1}$. This agrees with the analyses by King \latin{et al.},\cite{halas11angle} finding the scattering peak associated with the axial electric AE mode at a significantly shorter wavelength than that of the transverse electric (TE) modes.

To quantitatively analyze the modal composition, we extracted the electric field values distributed across the cap surface from numerical simulations and performed decomposition into spherical harmonics. Our analysis reveals that harmonics with $m=\pm1$ constitute the dominant contribution to the total field (Figure \ref{fig:basic-modes}C). Quantitatively, all other harmonic components collectively contributed less than 1\% to the total scattering power in the majority of cases, and never exceeded 3\% across all examined configurations. These findings provide strong validation for our theoretical model of the APTE mode.

Each distinct peak observed in the scattering spectra corresponds to a localized surface plasmon resonance at the respective excitation frequency. By analyzing the spherical harmonic decomposition, we assigned a characteristic wave number $k=\ell \cdot {R_\mathrm{JP}}^{-1}$ to each spectral peak, where $\ell$ represents the principal quantum number of the dominant harmonic $Y_\ell^m$. The relationship between LSP wave number and excitation frequency, as demonstrated in Figure \ref{fig:basic-modes}B, exhibits excellent agreement with the established plasmonic dispersion relation for interfacial systems\cite{BohrenHuffman,MaierPlasmonics,barnesSurfacePlasmonSubwavelength2003}

$$ \vert k_\mathrm{SP} \vert = \frac{\omega}{c} \sqrt{ \frac{\epsilon_1 \cdot \epsilon_2}{\epsilon_1 + \epsilon_2} } $$

which is valid for planar interfaces between materials with permittivities $\epsilon_1$ and $\epsilon_2$ and used as an approximation here for the curved interface.

These multimodal excitations match our models for the transverse-propagating modes and explain why peaks in the side-on spectrum are spaced more closely and are less prominent. The degeneracy of the APTE mode explains the difference in magnitude between the transverse-illuminated and axially-illuminated scattering cross-sections, the latter being approximately twice as large. This, again, matches the observations of first-order resonances by King \latin{et al.}\cite{halas11angle}

Surface roughness from thermal evaporation introduces scattering centers that limit surface plasmon propagation. When the scattering length becomes comparable to the mode wavelength, coherent standing-wave formation is disrupted, leading to broadening and intensity reduction. We suppose that modes with circumferential wavelengths $\lambda_{\text{circ}} > L_{\text{scat}}$ experience significant damping, providing a physical explanation for the observed spectral intensity drop-off at large wavelengths that is absent in our smooth-surface simulations.

\section{Conclusion \& Outlook}

In this work, we established Fourier Plane Tomographic Spectroscopy as a powerful analytical technique for characterizing orientation-dependent optical properties of individual micrometre-sized plasmonic Janus particles. This methodology, integrating dark-field microscopy with angle-resolved spectroscopy, enables simultaneous mapping of wavelength-dependent scattering across multiple illumination and detection angles. Our experimental measurements demonstrate quantitative agreement with finite-element simulations, validating both our measurement approach and theoretical framework.

Our spectroscopic analysis reveals that the structural asymmetry of pJPs induces distinct splitting of surface plasmon modes, resulting in unique spectral signatures that directly correlate with particle orientation. We observed characteristic features—including a distinct shoulder at 550 nm and multiple resonance peaks extending into the near-infrared—that clearly differentiate these asymmetric structures from isotropic gold spheres. Our mathematical framework based on spherical harmonic decomposition quantitatively explains these spectral features by establishing selection rules for surface plasmon modes on the curved metallic interface. This approach successfully identifies axial-propagating transverse-electric, transverse-propagating transverse-electric, and transverse-propagating axial-electric modes that dominate the scattering response.

The mode-specific angular distribution patterns we observed offer new insights into the complex interactions between polarized light and asymmetric plasmonic structures. By quantitatively correlating spectral features with specific surface plasmon modes, we have established a foundation for determining pJP orientation from scattering spectra, which could significantly enhance their utility in active matter research and self-propelled particle tracking. Additionally, our comprehensive characterization of orientation-dependent plasmonic responses provides critical design parameters for engineering optical manipulation strategies and sensing platforms based on these asymmetric structures.

The analytical approach using spherical harmonic decomposition developed here provides a robust framework for investigating complex light-matter interactions in asymmetric plasmonic nanostructures with applications ranging from nanophotonics to biomolecular sensing. Future refinements could include real-time spectral analysis for dynamic orientation tracking and the extension of our mathematical framework to complex geometries beyond hemispherical caps. The quantitative relationships we established between structural asymmetry and plasmonic mode splitting offer valuable design principles for engineering orientation-sensitive plasmonic responses in next-generation metamaterials and active colloidal systems.

\section{Methods}

\paragraph*{Sample Preparation}
The particles under investigation were plasmonic Janus particles consisting of spherical polystyrene (PS) cores of $1\,\si{\micro\metre}$ diameter hemispherically coated with gold. The gold coating was deposited to a thickness of $50\, \si{\nano\metre}$ with an intermediate $5\,\si{\nano\metre}$ chromium adhesion layer.
Samples were prepared by first depositing a volume of $30\, \si{\micro\litre}$ of the pJP suspension onto a glass cover slip. After allowing 10 minutes for sedimentation, the solvent was removed via nitrogen gas flow, resulting in pJPs immobilized on the cover slip surface. A second cover slip was placed on top with $1.5\, \si{\micro\litre}$ of immersion oil between the two cover slips. This sandwich configuration provided a homogeneous refractive index environment around the particles (see Fig. \ref{fig:setup}D).

\paragraph*{Experimental Setup}
The optical setup depicted in Fig. \ref{fig:setup} was constructed around an Olympus IX71 microscope platform. The numerical aperture of the dark-field condenser (Olympus U-DCW) ranged from 1.2 to 1.4, confining illumination to within 52.6° and 68.0° with respect to the optical axis. Aperture B1, constraining the illumination angle in-plane, was custom-manufactured with a slit width of 1 mm. Scattered light was collected using an Olympus UPlanFL N 100× objective with adjustable back aperture from 0.6 to 1.3 numerical aperture. All other lenses in the optical path were achromatic doublets with transmission spectral ranges from 400 to 1100 nm (Thorlabs AC-series). The spatial filter (B4) employed a pinhole with 0.9 mm diameter. The spectrograph consisted of a blazed transmission grating (Thorlabs GTI25-03A) mounted in front of an sCMOS camera sensor (PCO Edge 4.2), positioned behind an adjustable slit (Thorlabs VA100).

Switching between Fourier plane and real-space imaging was facilitated by translating the lens L2 between its two calibrated positions along the optical axis. While real-space imaging was used to select individual particles and acquire scattering spectra without angular resolution, back focal plane (BFP) imaging provided spectrally resolved Fourier-space scattering distributions. To acquire a 3D, angularly and spectrally resolved dataset, the slit B5 was translated across the BFP image, enabling sequential capture of vertical lines of spectrally dispersed scattering information. This was repeated systematically, for multiple fixed orientations of aperture B1 to obtain complete multidimensional scattering maps.

For angular and spectral calibration, the objective's back aperture (B3) was fully opened to its maximum numerical aperture of 1.3, allowing direct transmission of light from the dark-field illumination pathway into the imaging system. The image of the illumination pattern in the BFP was used to calculate the transformation from pixel coordinates to steric scattering angles. Meanwhile, the spectrally dispersed illumination pattern was used to determine the spectral response function of the setup, which is presented in Figure \ref{fig:setup}C. An optional long-pass filter could be used to red-shift the point of overlap between the first and second interference orders. Measurements with and without the filter were merged into spectroscopic datasets covering the full spectral range of the setup.

\paragraph*{Finite Element Simulations}
Numerical simulations of the LMI of an individual pJP were performed using COMSOL Multiphysics 6.1. The pJP's PS core was modelled as a sphere with diametre \mbox{1 µm} and refractive index 1.58. Its cap was represented by a half-ellipsoidal shell around the core, with a thickness of \mbox{50 nm} at the apex and \mbox{10 nm} at the rim. For the refractive index of gold, we used the values reported by Johnson and Christy.\cite{Johnson1972} This assembly was surrounded by an ambient medium with a constant refractive index of 1.51.

The simulations yielded numerical solutions for the scattered light field, given an incident plane wave. From the scattered field, optical cross-sections, far-field scattering intensity distributions, and point-wise solutions for the electric field on the particle surface were calculated.


\section*{Acknowledgements}
We thank A. M. Anton for insightful discussions. This measure is co-financed with tax revenue on the basis of the budget adopted by the Saxon State Parliament.

\section*{Author Contributions}
F.C. and F.H.P. designed the experiments; F.H.P. constructed the optical setup, performed the experiments, implemented the simulations and conducted the data analysis; F.H.P. wrote the manuscript; F.C. commented on the manuscript. All authors revised the manuscript.

\section*{Competing Interests}
The authors have no competing interests to declare.

\bibliography{References}

\end{document}